\documentclass[letterpaper]{article}

\usepackage[T1]{fontenc}

\usepackage{geometry}
\usepackage[utf8]{inputenc}  % Handles Unicode characters safely
\usepackage{amsmath}          % For math commands
\usepackage{amssymb}          % For \square

\usepackage{biblatex}
\usepackage{graphicx}
\usepackage{float}
\usepackage{subcaption}
\usepackage{siunitx}
\usepackage{dcolumn}
\usepackage{bm}
\usepackage{graphicx}    % for figures
\usepackage{float}       % for [H]
\usepackage{amsmath}

\usepackage{graphicx}
\usepackage{float}
\newfloat{scheme}{htbp}{los}
\floatname{scheme}{Scheme}
\floatname{chart}{Chart}
\newfloat{graph}{htbp}{loh}

\usepackage{chemformula} % Formulas using \ch{}
\usepackage[version = 4]{mhchem} % Formulas using \ce{}

\usepackage{authblk}

\author[1]{Francesca Incalza}
\author[1]{Matteo Castellani}
\author[1]{Dip Joti Paul}
\author[1]{Alejandro Simon}
\author[1]{Emma Batson}
\author[1]{Davide Mondin}
\author[1]{Owen Medeiros}
\author[1]{Karl K. Berggren}

\affil[1]{Department of Electrical Engineering and Computer Science, Massachusetts Institute of Technology, Cambridge, MA, USA}
\date{*Email: incal723@mit.edu}

\title{Fast-Recovery Epitaxial NbN Superconducting Nanowire Single-Photon Detectors with Saturated Efficiency at 1550 nm in Liquid Helium}

\begin{document}
%\setstretch{2.5}

\maketitle

\begin{abstract}

Achieving both high internal efficiency and fast reset times at elevated temperatures remains challenging due to limited understanding of how film properties govern SNSPD performance. We demonstrate that epitaxial NbN films on sapphire enable simultaneous high efficiency and rapid response. We fabricate and characterize SNSPDs based on these films deposited via DC magnetron sputtering on c-cut sapphire. High-quality epitaxial growth preserves a low electron diffusion coefficient and promotes strong electron–phonon coupling, yielding a high critical temperature and efficient hotspot formation in the dirty limit. X-ray diffraction and transmission electron microscopy confirm epitaxial alignment and lattice order. Nanowires of \SI{20}{\nano\meter} width exhibit saturated internal efficiency at \SI{1550}{\nano\meter} wavelength and short reset times at \SI{4.2}{\kelvin}, enabled by lattice matching and high thermal conductance of the sapphire interface. Ab initio modeling reproduces photon count rates, linking device performance quantitatively to film properties such as diffusivity and electron–phonon coupling.
\end{abstract}


Thin, high-quality superconducting films are essential for the advancement of quantum technologies, enabling the development of high-performance devices such as superconducting nanowire single-photon detectors (SNSPDs)~\cite{goltsman_picosecond_2001, esmaeil_zadeh_superconducting_2021} and superconducting electronics~\cite{buzzi_nanocryotron_2023, castellani_nanocryotron_2024, foster_superconducting_2023, castellani_superconducting_2025, medeiros_scalable_2025}. SNSPDs are a key technology for applications ranging from integrated quantum photonic circuits~\cite{colangelo_molybdenum_2024} and biological imaging~\cite{lin_surface_2022} to deep-space optical communication~\cite{wollman_recent_2021}, owing to their high detection efficiency (DE)~\cite{natarajan_superconducting_2012}, ultra-low dark count rates~\cite{shibata_ultimate_2015}, and high temporal resolution~\cite{korzh_demonstration_2020, autebert_direct_2020}.
These performance metrics are inherently tied to the material properties of the superconducting film~\cite{holzman_superconducting_2019} and strongly dependent on substrate choice~\cite{steinhauer_nbtin_2020}, film thickness~\cite{hofherr_intrinsic_2010}, and nanowire geometry~\cite{natarajan_superconducting_2012}. As a result, optimizing the material through the deposition process, whether the method is sputtering, atomic layer deposition, or molecular beam epitaxy, is essential for achieving reproducible high-performance devices.\\
SNSPDs are typically fabricated from polycrystalline nitride superconductors such as niobium nitride (NbN)~\cite{zhang_nbn_2017} and niobium titanium nitride (NbTiN)~\cite{esmaeil_zadeh_single-photon_2017} or from amorphous materials such as molybdenum silicide (MoSi)~\cite{verma_high-efficiency_2015} and tungsten silicide (WSi)~\cite{verma_high-efficiency_2014}. Amorphous material-based SNSPDs exhibit uniform films across large substrates and system detection efficiencies up to $\sim$93\% at 1550~nm~\cite{verma_high-efficiency_2015,verma_high-efficiency_2014}, but they generally require temperatures below 1~K and display high kinetic inductance ($\sim$1~$\mu$H), limiting their maximum count rates.\\
In contrast, polycrystalline nitride SNSPDs benefit from higher critical temperatures (4-5~K), which enable operation at more accessible cryogenic conditions, as well as superior timing performance with jitter below 3~ps~\cite{korzh_demonstration_2020}. However, their polycrystalline nature leads to non-uniform critical currents and the presence of grain boundaries, which can reduce fabrication yield and complicate scaling to large detector arrays. In particular, variations in the local critical current density make it increasingly difficult to uniformly bias the nanowire at elevated operating temperatures, such as 4.2~K (liquid helium), limiting the ability of the detector to reach saturation across the entire active area and thereby reducing system detection efficiency~\cite{kerman_constriction-limited_2007}.\\
Despite the appeal of high-$T_\mathrm{c}$ superconductors such as MgB\textsubscript{2}, which can enable single-photon detection up to 20 K~\cite{batson_effects_2024,charaev_single-photon_2024,incalza_effects_2023}, their implementation in SNSPDs remains challenging. These devices typically show relatively low detection efficiencies and count rates~\cite{cherednichenko_low_2021}. This limitation is likely connected to the epitaxial nature of MgB\textsubscript{2} films, where the well-ordered lattice favors high $T_\mathrm{c}$ but does not necessarily optimize the microscopic conditions for efficient photon detection.

NbN remains one of the most widely adopted materials due to its fast response time~\cite{korzh_demonstration_2020} and higher superconducting critical temperature compared to amorphous alternatives.
Epitaxial NbN has emerged as a promising route to reduce grain boundaries and defect density, yielding films with improved uniformity, higher critical temperatures, and better device performance. Previous works demonstrated excellent SNSPD performance~\cite{marsili_high_2008, marsili_single-photon_2011}, though the epitaxial nature of these films was not explicitly confirmed. NbN epitaxy has also been achieved on GaAs~\cite{reithmaier_optimisation_2013}, 3C–SiC~\cite{dochev_growth_2011}, and various sapphire orientations~\cite{espiau_de_lamaestre_microstructure_2007, villegier_epitaxial_2009, lamaestre_high_2008}. Films deposited with techniques including pulsed laser deposition~\cite{roch_substrate_2021}, PEALD~\cite{shibalov_ultrathin_2023}, HTCVD~\cite{hazra_superconducting_2016}, and MBE~\cite{wright_growth_2023} have reached $T_\mathrm{c}$ up to 15.5~K on sapphire, but sputtering remains attractive as it offers faster deposition rates and lower operational costs. Indeed, sputter-deposited epitaxial NbN films on AlN-buffered sapphire have achieved $T_\mathrm{c}$ up to 11.2~K (5~nm thick)~\cite{wei_ultrathin_2021}, and films on $c$-cut sapphire have reached 14.8~K (5~nm)~\cite{haberkorn_probing_2024}.\\
High-quality epitaxial NbN SNSPDs have been realized on AlN-buffered sapphire using MBE~\cite{cheng_epitaxial_2020}, AlN-buffered Si~\cite{iovan_nbn_2023}, diamond~\cite{atikian_superconducting_2014}, and MgO~\cite{marsili_high_2008,miki_nbn_2007}. Despite these advances, a central challenge for NbN SNSPDs remains: combining a high superconducting critical temperature ($T_\mathrm{c}$) with the microscopic material properties that favor efficient single-photon detection. The underlying trade-offs are not yet fully understood, because the conditions that stabilize a high $T_\mathrm{c}$, such as crystalline order and strong electron-phonon coupling, do not always coincide with those that maximize internal detection efficiency, which typically benefits from enhanced quasiparticle scattering and more uniform current redistribution. Indeed, although epitaxial NbN can exhibit enhanced $T_\mathrm{c}$, previous reports~\cite{cheng_epitaxial_2020} showed that saturation efficiency often remained below that of optimized polycrystalline devices ($\sim$70–80\%). In that work, the films were nearly stoichiometric and had very low defect density, resulting in high electron diffusivity. While this high diffusivity contributed to the elevated $T_\mathrm{c}$, it also reduced hotspot confinement, limiting internal detection efficiency. In contrast, in this study, we demonstrate that high $T_\mathrm{c}$ and high internal efficiency are not mutually exclusive: by carefully controlling nanowire geometry, defect density, and electron-phonon interactions in our epitaxial NbN films, we achieve both.

Epitaxial NbN SNSPDs generally exhibit reduced interfacial stress and strain, which helps preserve a high $T_\mathrm{c}$ and maintain uniform superconducting properties across the nanowire. In this work, we intentionally engineered the film deposition to introduce a controlled density of atomic defects while maintaining high $T_\mathrm{c}$. This optimization ensures that the films remain in the dirty-limit regime ($l \ll \xi$, where $l$ is the electron mean free path and $\xi$ the coherence length), where quasiparticles undergo frequent scattering from residual atomic defects~\cite{tsuneto_dirty_1962}, a key requirement for achieving saturated internal detection efficiency. Improved lattice matching also modifies phonon transmission, affecting thermal coupling and energy relaxation following photon absorption. The coexistence of scattering and strong coupling accelerates quasiparticle trapping and confinement, enabling saturated internal efficiency and fast recovery dynamics~\cite{ilin_picosecond_2000}. Notably, films grown on sapphire exhibit especially fast reset times, often attributed to efficient phonon transmission between the epitaxial NbN film and the sapphire substrate, which enables cooling and rapid recovery after photon absorption~\cite{kerman_kinetic-inductance-limited_2006,zhang_response_2003,pearlman_gigahertz_2005}.\\
\begin{figure}[H]
\includegraphics[width=\linewidth]{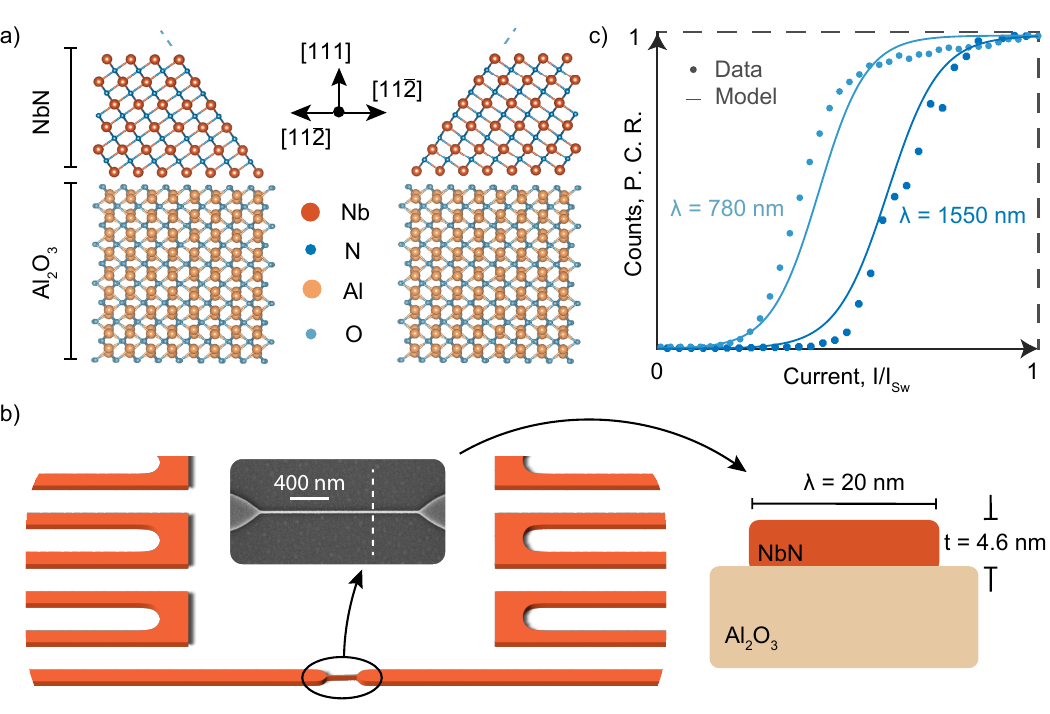}
\caption{\raggedright % This aligns the text
\label{fig:Total} Structural and device characteristics of NbN nanowire superconducting single-photon detectors (SNSPDs). (a) Atomic model of epitaxial NbN on Al$_2$O$_3$, showing NbN with a [111] orientation and the Al$_2$O$_3$ atomic lattice~\cite{momma_vesta_2011}. Two twin-related growth variants linked by a $180^\circ$ rotation about $\langle111\rangle$ are observed, with orientation relationships (111)NbN$\parallel$(0006)Al$_2$O$_3$ and $\pm\langle\bar{1}10\rangle$NbN$\parallel\langle10\bar{1}0\rangle$Al$_2$O$_3$. (b) SNSPD layout and nanowire stack: left, schematic of the meander device with an SEM image of the central nanowire; right, layer stack of a \SI{20}{nm}-wide NbN nanowire (thickness \SI{4.6}{nm}) on Al$_2$O$_3$. (c) Normalized photon count rate (PCR) versus normalized bias current $I/I_{\mathrm{sw}}$ for incident wavelengths \SI{780}{nm} and \SI{1550}{nm} for a 20~nm-wide nanowire. Dots: experimental data; solid lines: \textit{ab initio} model fits; measurements taken at \SI{4.2}{K}, showing saturation of detection efficiency at both wavelengths.
}
\end{figure}
In this paper, we show that SNSPD performance is strongly determined by material properties such as electron diffusivity, superconducting critical temperature, and sheet resistance. Specifically, lower diffusivity and controlled defect density enhance internal detection efficiency by confining photon-induced hotspots, high $T_\mathrm{c}$ enables operation at elevated temperatures, and the choice of substrate contributes to fast reset times. Figure 1 provides an overview of this approach. We deposited epitaxial NbN thin films on c-cut sapphire substrates through DC reactive ion sputtering. By optimizing the deposition process, we identified films that combined epitaxial order with the low electron diffusivity typical of polycrystalline films, achieving both a high superconducting critical temperature ($T_\mathrm{c}$) and properties favorable for efficient hotspot formation. To understand how epitaxial growth influences the superconducting behavior of NbN, we performed a characterization of the films structural, morphological, and superconducting properties. We began by directly probing the structural quality, investigating the crystallinity and epitaxial alignment. Our films exhibit high epitaxy, with a superconducting critical temperature reaching up to 15.5~K for a 20~nm thick film. Structural analysis revealed an epitaxial relationship as shown in figure~\ref{fig:Total}a characterized by two twin-related growth variants connected by a 180$^\circ$ rotation around the $\langle111\rangle$ directions, with orientation relationships $ (111) $NbN$\parallel$(0006)Al$_2$O$_3$ and $\pm\langle\bar{1}10\rangle$NbN$\parallel$$\langle10\bar{1}0\rangle$~Al$_2$O$_3$ (see Supporting Information for more information on the epitaxial relation). Importantly, we also demonstrate that these epitaxial films are in the dirty-limit regime, with a measured electron diffusivity of approximately 0.5~cm$^2$/s. We fabricated SNSPDs using epitaxial NbN thin films grown by the process mentioned above. The resulting devices, consisting of 20~nm-wide nanowires with a thickness of 4.6~nm (Figure~\ref{fig:Total}b), demonstrate saturated internal detection efficiency at 780 and 1550~nm wavelengths, even when operated at 4.2~K (Figure~\ref{fig:Total}c). Furthermore, the detectors exhibit fast reset times on the order of hundreds of picoseconds, attributed to the epitaxial interface with the highly thermally conductive sapphire substrate. Finally, we applied an \textit{ab initio} model~\cite{simon_ab_2025}, informed by detailed material characterization, to quantitatively reproduce the experimental photon count rate curves and to discuss the model's potential to improve SNSPD design. 
\begin{figure}[H]
\centering
\includegraphics[width=\linewidth]{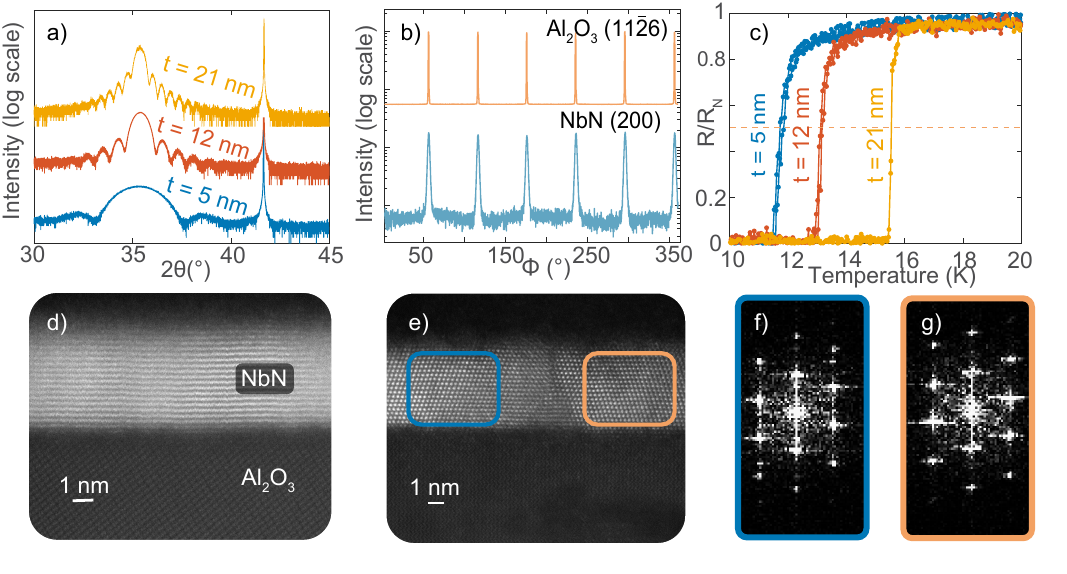}
\caption{\raggedright
\label{fig:fig2} Structural and Superconducting Characterization of Epitaxial NbN Thin Films. (a) X-ray diffraction (XRD) \( \theta \)-2\( \theta \) scans (log scale) for NbN films of thickness 5 nm, 12 nm, and 21 nm, exhibiting clear interference fringes and a sharp (111) NbN peak, confirming high crystalline quality. (b) Azimuthal (\( \phi \)) scans of the NbN (200) and sapphire (11$\bar{2}$6) reflections, both displaying sixfold symmetry. The sixfold pattern in NbN arises from the presence of rotational twin variants induced by epitaxial growth on the hexagonal sapphire substrate, consistent with cube-on-hexagonal epitaxy. (c) Normalized resistance vs.\ temperature curves for the three film thicknesses, illustrating sharp superconducting transitions and thickness-dependent \( T_c \) values. (d) Cross-sectional TEM image of a 4.5~nm NbN thin film showing the out-of-plane epitaxial relationship, where the NbN [111] direction is aligned with the sapphire [0001] axis. (e) TEM image highlighting well-defined twin domains (indicated by the rectangle), which are characteristic of the epitaxial growth on sapphire. (f–g) Fast Fourier Transforms (FFTs) extracted from the boxed region in (e), showing mirrored diffraction patterns that confirm the presence of twin domains.
}
\end{figure}
Figure~\ref{fig:fig2}a shows high-resolution X-ray diffraction (XRD) $\theta$-2$\theta$ scans (log scale) for representative 5~nm, 12~nm, and 21~nm thick films. The strong NbN (111) diffraction peak in all three samples indicates preferential out-of-plane orientation and confirms face-centered cubic NbN with no secondary phases. Pronounced interference fringes across all thicknesses, including the 5~nm film, indicate good surface smoothness and high uniformity. The FWHM of the (111) peak shows a slight increase with decreasing thickness, consistent with strain relaxation effects in ultrathin NbN films~\cite{farhadizadeh_structure_2025}. Applying the Scherrer equation~\cite{uvarov_metrological_2007} to the (111) peak width, the vertical out-of-plane crystallite size was estimated to be comparable to the film thickness measured by ellipsometry (find details in the Supporting Information). Figure~\ref{fig:fig2}b shows azimuthal ($\phi$) scans of the NbN (200) and sapphire (11$\bar{2}$6) reflections. Both exhibit sixfold symmetry due to rotational twin domains rotated by 90$^\circ$ out-of-plane. This symmetry indicates cube-on-hexagonal epitaxy, with NbN [111] aligned to sapphire [0001] and in-plane relationship [200]$_\text{NbN} \parallel$ [11$\bar{2}$6]$_\text{Al2O3}$. Transmission electron microscopy (TEM) images in Figure~\ref{fig:fig2}d and ~\ref{fig:fig2}e further confirm epitaxial [111] alignment, reveal twin domains, and show well-defined grains validated by the mirrored Fast Fourier Transform (FFT) patterns in ~\ref{fig:fig2}f and ~\ref{fig:fig2}g. Such rotational twin boundaries, characteristic of cube-on-hexagonal epitaxy, may locally disrupt in-plane uniformity and act as scattering centers for electrons and phonons; however, they are generally more coherent than random grain boundaries, with lower defect densities and reduced strain, making them less likely to strongly suppress local critical current~\cite{lin_characterization_2013}. In addition, these twins can also serve as vortex pinning centers, influencing vortex dynamics~\cite{lin_characterization_2013}, and potentially affect hotspot nucleation and detection efficiency. Overall, the high structural quality achieved through epitaxy translates into enhanced superconducting behavior, as reflected in the high critical temperatures of these films. 

Figure~\ref{fig:fig2}c shows the normalized resistance versus temperature for NbN films with thicknesses of 5~nm, 12~nm, and 21~nm. The critical temperature $T_c$, defined at 50\% of the resistive transition, increases with thickness from 11.8~K to 15.6~K, approaching the bulk value of 16.6~K~\cite{zhang_superconductivity_2019}. The superconducting transition widths decrease from 1.66 to 0.33~K as the film thickness increases, indicating that thicker films are more uniform~\cite{baeva_natural_2024}. Additional experimental details are provided in the Supporting Information.
The residual resistance ratio (RRR = $R(300~\mathrm{K})/R(20~\mathrm{K})$) increases from 2.31 to 13.50 with thickness increasing from 5~nm to 21~nm, reflecting improved metallicity and structural quality~\cite{roach_nbn_2012}. Additional structural and surface characterization, including AFM, EBSD, and XPS, is provided in the Supporting Information.
To further explore the influence of of the NbN film microstructure on superconductivity, we measured the critical temperature $T_c$ under different perpendicular magnetic fields (additional information is provided in the Supporting Information). From these measurements, the upper critical field at zero temperature for a \SI{5}{\nano\metre} thick film, $\mu_0 H_{c2}(0)$, was estimated using the Werthamer–Helfand–Hohenberg (WHH)~\cite{werthamer_temperature_1966} approximation in the dirty limit as $\mu_0 H_{c2}(0) = 0.69\, T_c \left| \frac{dH_{c2}}{dT} \right|_{T \to T_c}  \approx 17~\mathrm{T}$. The superconducting coherence length was estimated to be $\xi(0) = \sqrt{\frac{\Phi_0}{2 \pi B_{c2}(0)}}  \approx 4.4~\mathrm{nm}$, and the electron diffusivity is $D = \frac{4 k_B}{\pi e} \cdot \left| \frac{dB_{c2}}{dT} \right|^{-1} \approx 0.47~\mathrm{cm^2/s} = 47~\mathrm{nm^2/ps}$. This low diffusivity places the film in the dirty-limit regime, where the electron mean free path is much shorter than the coherence length ($l = 3D/v_F \sim 5$~\AA\ $\ll \xi \sim 4.4~\mathrm{nm}$), indicating that quasiparticles undergo frequent scattering from residual atomic defects and lattice imperfections.
\begin{figure}[H]
\centering
\includegraphics[width=0.5\linewidth]{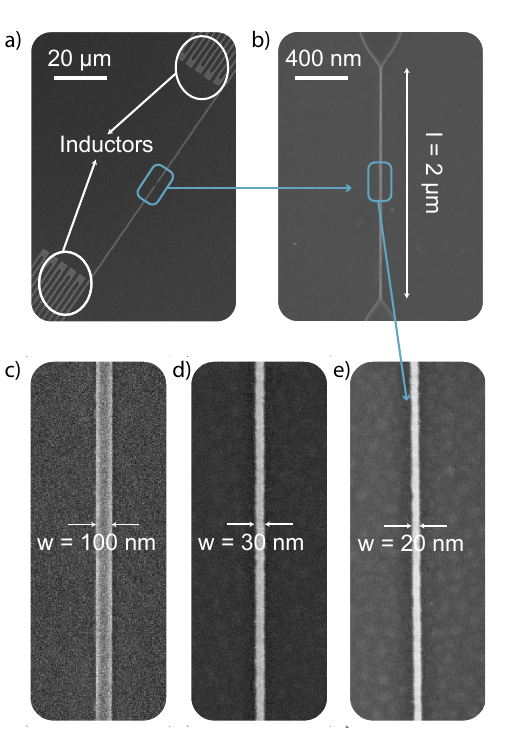}
\caption{ Nanofabrication and dimensional characterization of the NbN nanowire device. (a) Scanning electron micrograph (SEM) of the overall device layout. The structure comprises NbN nanowires of different widths integrated with inductors. Each \SI{2}{\micro\meter}-long straight nanowire is serially connected to meandered sections formed by \SI{500}{nm}-wide wires acting as inductors. (b) Magnified SEM image of the central nanowire region, where the narrowest section extends over approximately \SI{2}{\micro\meter}. (c, d, e) High-resolution SEM micrographs illustrating the precision of the electron-beam lithography and etching processes used to define the nanowire width. The three images correspond to nanowires fabricated with nominal widths of \SI{100}{nm}, \SI{30}{nm}, and \SI{20}{nm}, respectively.
}
\label{fig:figure3}
\end{figure}
We fabricated nanowires from epitaxial NbN with different widths to investigate how geometrical confinement and the underlying microstructure influence device performance. The fabrication process employed high-resolution electron-beam lithography based on a negative-tone resist and development in 25\% tetramethylammonium hydroxide (TMAH) at room temperature, followed by reactive ion etching to transfer the pattern into the NbN layer. See Supporting Information for additional fabrication and process parameters. Figure~\ref{fig:figure3} shows a scanning electron microscopy (SEM) image of a representative 20-nm-wide nanowire patterned with integrated inductors to enable self-reset. All devices consist of \SI{2}{\micro\meter}-long straight nanowires serially connected to long meandered wires, each 500~nm wide. The nanowire widths were extracted from the SEM images as described in the Supporting Information. The room-temperature sheet resistance of the film prior to fabrication is  491~$\Omega$/sq and the film thickness of 4.6~nm.

\begin{figure}[H]
\centering
\includegraphics[width=\linewidth]{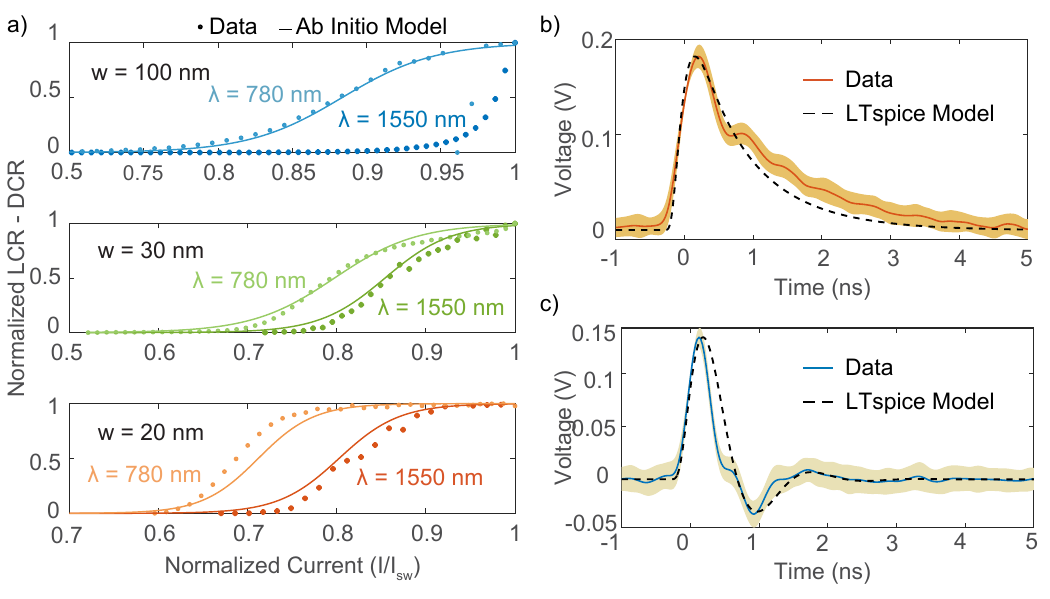}
\caption{\raggedright
\label{fig:pcr} Photon detection performance and electrical response of nanowire detectors. 
(a) Normalized photon count rate (PCR, light minus dark) versus normalized bias current ($I/I_{\mathrm{sw}}$) for nanowires of widths \SI{20}{nm}, \SI{30}{nm}, and \SI{100}{nm}. Data points correspond to incident photon wavelengths of \SI{780}{nm} and \SI{1550}{nm} measured at \SI{4.2}{K}. The two zero-valued points in the 100 nm curve at 1550 nm are attributed to a transient acquisition artifact, most likely caused by counter saturation or a brief timeout. Solid lines represent \textit{ab initio} model fits, showing that narrower nanowires (e.g., \SI{20}{nm}) reach unity detection efficiency at lower bias currents. (b) Temporal response obtained from 100 individual \SI{100}{nm} superconducting nanowire traces. The mean measured voltage output (orange solid line) and theoretical model fit (dashed black line) are shown as a function of time. The shaded orange area represents $\pm$1 standard deviation. (c) Temporal response of the same device as in (b) with an additional series resistor, measured over 100 traces and approaching the estimated $L/R$ bandwidth limit of \SI{120}{ps}. The mean measured voltage (blue solid line) and model fit (dashed black line) illustrate the fast reset dynamics enabled by the resistor, with the shaded yellow area indicating $\pm$1 standard deviation.
}
\end{figure}

Figure~\ref{fig:pcr}a presents the normalized photon count rate (PCR) versus normalized bias current for detectors illuminated at 1550~nm and 780~nm. The switching currents are 15~\textmu A (100~nm), 3.3~\textmu A (30~nm), and 2.5~\textmu A (20~nm), corresponding to a critical current density of $\sim$4~MA/cm\(^2\). The 20~nm and 30~nm devices display clear saturation at 780~nm and a more gradual slope at 1550~nm, whereas the 100~nm devices show no saturation, particularly at longer wavelengths. 

This behavior is largely geometric: with a coherence length of 5.74~nm, the Likharev criterion~\cite{likharev_superconducting_1979} sets the minimum wire width for vortex formation at $4.4\xi \approx 25$~nm. Vortices are therefore forbidden in the 20~nm wire and would occupy a large fraction of the 30~nm wire, allowing a photon-induced hotspot to switch the entire width almost instantaneously. In contrast, a 100~nm wire generates a vortex-antivortex pair~\cite{simon_ab_2025}, and detection occurs only if the vortices reach the edges before the hotspot dissipates. Narrow wires thus leverage the benefits of epitaxial films, high $T_c$ and fast reset times, without efficiency loss from rapid quasiparticle cooling. Furthermore, having a film thickness smaller than the superconducting coherence length ensures that the suppression of the order parameter is nearly uniform across the film thickness. This dimensional constraint enhances quasiparticle localization within the hotspot, reducing the energy required to initiate a detection event and enabling efficient photon detection at lower bias currents.

The observed differences in performance can be further understood by considering the electronic properties of the epitaxial NbN films. While scattering from nitrogen vacancies and other defects shortens the electron mean free path, promoting uniform hotspot formation and rapid quasiparticle relaxation, the high-quality epitaxial lattice preserves strong electron–phonon coupling, maintaining a high $T_c$. Together, these features enable narrow SNSPDs to achieve high internal efficiency even at elevated operating temperatures. Scattering in the films mainly occurs at twin boundaries and defects, but these features do not affect the film quality. While electrons are scattered by Ångström-scale imperfections, the longer-wavelength phonons (on the order of nanometers in NbN) remain largely unaffected, ensuring robust electron–phonon coupling and preserving the epitaxial order.

To obtain further insight into the PCR behavior and guide device optimization, we complemented the experiments with a first-principles model combining \textit{ab initio} kinetic equations and time-dependent Ginzburg--Landau (TDGL) simulations~\cite{simon_ab_2025}. Using our measurements for $D$, $R_{s}$, $T_{\mathrm{c}}$, and $d$ from our detailed epitaxial NbN characterization as inputs into the model, we predicted the theoretical saturation current for detection for the different detector geometries that were fabricated. This yielded a model that closely matches the experimental observations, as seen in Figure~\ref{fig:pcr}a, where the solid lines show the results of this combined modeling and the 50\% point in the PCR curve is the saturation current. In these simulations, the width of the PCR curves and the constriction factor was fit to account for Fano fluctuations, thermal fluctuations, and other stochastic processes during photon detection that are not accounted for in the calculation of the saturation current. We did not attempt to calculate the saturation current at 1550~nm for the 100~nm wide wire since the wire did not saturate at that wavelength, as it was too constricted to reach saturation. The agreement between theory and experiment allows us to establish a feedback loop between theory and experiment to optimize device performance. This agreement further confirms that the high detection efficiency and performance arise directly from the optimized geometry and intrinsic material properties of the fabricated SNSPDs. Further details on the model are contained in the Supporting Information. 

To demonstrate the fast response time of our SNSPDs, we reduced the $L/R$ time constant by placing a surface-mount resistor at 4.2~K in series between the 100 nm-wide detector and the 50~$\Omega$ matched amplifier. We selected the wider detector for this measurement because its larger critical current yields a higher output pulse amplitude, providing an improved signal-to-noise ratio. We repeated the experiment several times, increasing the resistor value each time until reaching the estimated $L/R$ bandwidth limit of 120~ps, calculated from the kinetic inductance of the nanowire and the nominal load resistance~\cite{wang_controlling_2025}. Figures~\ref{fig:pcr}b and c show the temporal responses of the detectors with and without the added resistor, respectively. The shaded area represents $\pm 1$ standard deviation over 100 individual superconducting nanowire traces, while the solid line shows the mean pulse and the dashed line represents the theoretical model (more information can be found in the Supporting Information). The longer measured recovery time compared to the designed $L/R$ is attributed to finite bandwidth of the readout electronics, impedance mismatches, and the influence of parasitic capacitance in the readout chain. Electrical modeling of our system shows that introducing a 1 pF capacitance to ground between the SNSPD source and the load resistor reproduces the slight post-pulse ringing observed in Figure~\ref{fig:pcr}c. This parasitic capacitance becomes more apparent at higher load resistances, where the increased $R$ makes the circuit more sensitive to small capacitive effects. From a single experimental waveform, the electrical pulse in Figure~\ref{fig:pcr}c exhibits a full width at half maximum (FWHM) of approximately 400~ps, while the recovery time from 90\% to 10\% amplitude is about 550~ps. The measured recovery time was affected by the finite bandwidth of the readout electronics and impedance mismatches in the amplification chain. The device reliably detected photons and reset without latching. Figure~\ref{fig:pcr}b shows the same device without the additional load resistor, exhibiting a decay time constant (see Supporting Information for details) of 1.2~ns. Despite the short reset time, the maximum measured count rate was $\sim 7\times10^7$ counts per second, which we attribute primarily to limited photon flux and overall system detection efficiency. The corresponding photon count rate curve may be found in the Supporting Information. Future optimization of optical coupling and system efficiency could enable higher count rates, fully exploiting the fast reset time of the SNSPDs.

The rapid reset time achieved by increasing the load resistance and observed consistently across different wire widths, is primarily attributed to the combination of the high-quality epitaxial NbN film and the excellent thermal coupling to the sapphire substrate~\cite{berggren_superconducting_2018}. The epitaxial lattice ensures uniform superconducting properties with minimal local variations, while the sapphire provides a high thermal conductance pathway ($h_c \approx 9.3 \times 10^4~{\rm W m^{-2} K^{-1}}$, see Supporting Information for details), allowing the excess energy deposited by photon absorption to be efficiently dissipated. Together, these factors enable the quasiparticles and phonons generated in the hotspot to relax quickly, restoring the superconducting state and allowing the nanowire to detect subsequent photons with minimal dead time.

In conclusion, we have demonstrated that epitaxial NbN thin films grown on $c$-cut sapphire via reactive DC sputtering constitute a useful material platform for high-performance superconducting nanowire single-photon detectors. The high-quality epitaxial growth preserves strong electron-phonon coupling and ensures low electron diffusivity, resulting in a high critical temperature and efficient hotspot formation in the dirty-limit regime. SNSPDs fabricated from these films, with nanowire widths down to 20~nm, exhibit saturated internal detection efficiency at both 780~nm and 1550~nm wavelengths, even at 4.2~K, and feature extremely fast reset times attributed to the epitaxial lattice and the thermally conductive sapphire interface. These results establish that sputter-deposited epitaxial NbN on sapphire, grown under conditions that preserve epitaxial order while maintaining a controlled defect density, combines high $T_\mathrm{c}$, low electron diffusivity, and controlled microstructure to enable SNSPDs with high internal efficiency and fast recovery for well-confined wires. The experimental photon count rate curves are fitted to an \textit{ab initio} model, confirming that device performance is determined by the intrinsic properties of the epitaxial films. By directly linking electron diffusivity, critical temperature, and microstructural characteristics to device behavior, we show how intrinsic material properties govern hotspot formation and operational temperature, providing guidance for the design of SNSPDs across different superconducting platforms.

\section*{Acknowledgements}

We wish to acknowledge the support of J. Daley and C. Moorman for their assistance with SEM imaging, C. Settens for help with material characterization, L. Shaw for the help with XPS, M. Mondol for help with e-beam lithography and device fabrication and A. Buzzi for assistance with images.\\
We are also grateful to Rohit Prasankumar, Matthew Julian, Stu Wolf, Quanxi Jia, and Mukund Vengalattore for helpful discussions, as well as to Phillip Donald Keathley, Dongmin Kim and Ben Mazur for their insightful feedback on this manuscript. \\
TEM characterization was kindly performed by Eurofins EAG Laboratories, whose partial support we gratefully acknowledge.\\
This work was funded in part by the Defense Sciences Office (DSO) of the Defense Advanced Research Projects Agency (DARPA) (HR0011-24-9-0311). FI acknowledge support from the Frank Quick Fellowship, AS and EB acknowledge support from the NSF Graduate Research Fellowship Program (GRFP), DJP acknowledges support from the MathWorks Engineering Fellowship and OM acknowledges support from the NDSEG Fellowship.

%%%%%%%%%%%%%%%%%%%%%%%%%%%%%%%%%%%%%%%%%%%%%%%%%%%%%%%%%%%%%%%%%%%%%
%% If you are using classical BibTeX rather than biblatex,
%% remove the \printbibliography and uncomment the \bibliograpy one
%%%%%%%%%%%%%%%%%%%%%%%%%%%%%%%%%%%%%%%%%%%%%%%%%%%%%%%%%%%%%%%%%%%%%

%\newpage
\ref{supp}

\title{\LARGE \textbf{Supporting Information}}
\label{supp}
%\setstretch{2.5}
\renewcommand{\thesection}{S\arabic{section}}
\setcounter{section}{0}
\maketitle

\tableofcontents
\newpage

\section{Epitaxial relation of NbN on c-cut Al$_2$O$_3$}

\begin{figure}[h]
    \centering
    \includegraphics[width=\linewidth]{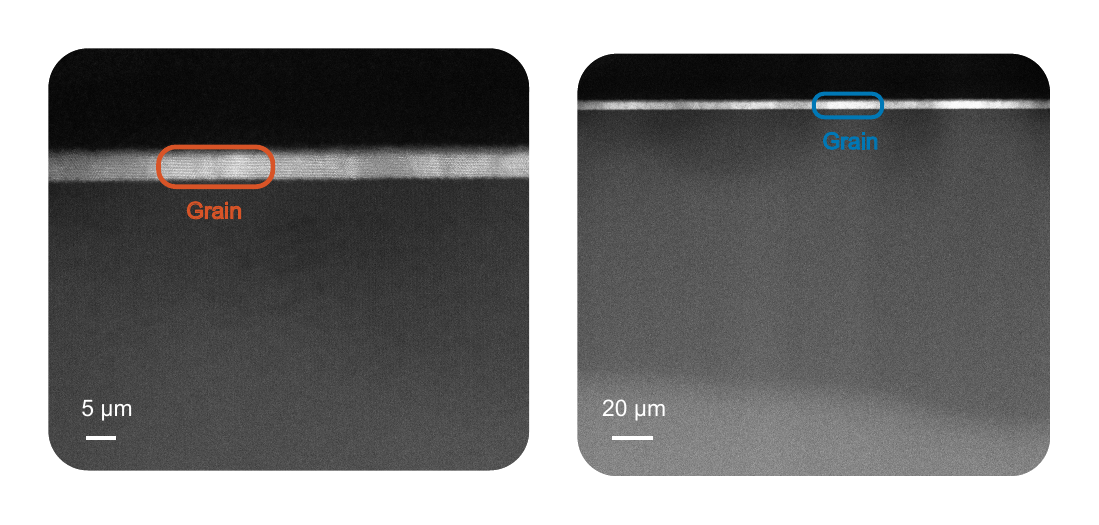} 
    \caption{Transmission electron microscopy image of epitaxial NbN on c-cut Al$_2$O$_3$. Twin domains are visible as regions with alternating contrast (grey variations). The lateral size of the twins is on the order of a few nanometers.}
    \label{fig:S5}
\end{figure}

The epitaxial growth of NbN on c-cut Al$_2$O$_3$ naturally leads to the formation of twin domains. This occurs because the cubic structure of NbN on the hexagonal sapphire lattice allows for two symmetrically equivalent orientation variants, related by a 180° rotation around the [111] zone axis. These twins typically have lateral dimensions on the order of a few nanometers, as observed in the TEM image shown in Fig.~\ref{fig:S5}. In this TEM, the changes in contrast of the grey regions allow us to identify the different twin domains.

The epitaxial relation can be rationalized using the coincidence site lattice rule~\cite{shibalov_ultrathin_2023}, which describes the alignment of two lattices when a subset of lattice points from each crystal coincide. In this framework, epitaxial growth is favored along directions where the lattices share a periodic coincidence, even if the overall lattice constants are different. For NbN on sapphire, the rule predicts:

\begin{equation}
4 \cdot \sqrt{3} \cdot \sqrt{2} \, a_{\rm NbN} = 24.856~\text{\AA}, \quad
3 \cdot \sqrt{3} \, a_{\rm Al_2O_3} = 24.728~\text{\AA},
\end{equation}

which corresponds to a lattice mismatch of less than 0.52\%. This low mismatch explains the observed epitaxial alignment and why twin domains form: the crystal accommodates the small residual strain by forming two equivalent orientations.

Additionally, from X-ray diffraction (XRD) measurements, the (111) peak of NbN was used to extract the interplanar spacing $d_{111}$. Using the cubic lattice geometry, the in-plane lattice parameter $a$ is related to $d_{111}$ by

\begin{equation}
a = d_{111} \cdot \sqrt{3} = 2.537~\text{\AA} \cdot \sqrt{3} \approx 4.39~\text{\AA}.
\end{equation}

This value is in good agreement with the expected bulk lattice parameter.

\section{Thickness-dependent structural and superconducting properties}

\begin{figure}[ht]
\centering
\includegraphics[width=0.5\linewidth]{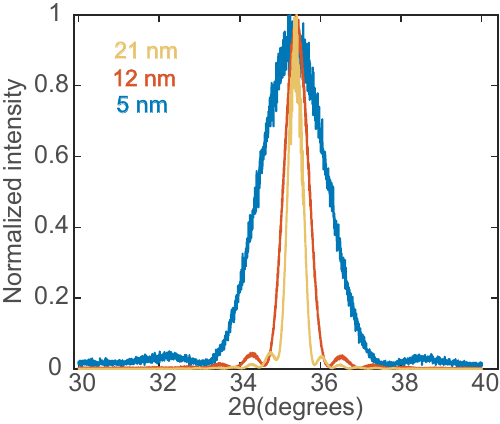}
\caption{\raggedright
\label{fig:XRD_PEAK}
Normalized X-ray diffraction (XRD) peak for NbN (111) plane for films of different thicknesses. The overlapping curves highlight differences in peak broadening associated with film thickness, indicating variations in crystallinity and strain.}
\end{figure}

The superconducting and structural properties of the epitaxial NbN films exhibit a clear dependence on film thickness, consistent with strain relaxation in thin films. X-ray diffraction measurements reveal that thicker films (Fig.~\ref{fig:XRD_PEAK}) have narrower rocking curves (FWHM $<$ 1$^\circ$ for 21\,nm), indicating low mosaic spread and out-of-plane alignment. The crystallite size \(D\) was estimated using the Scherrer formula~\cite{uvarov_metrological_2007}, 
\[
D = \frac{K \lambda}{\beta \cos\theta},
\] 
where \(K \approx 0.9\) is the shape factor, \(\lambda\) is the X-ray wavelength, \(\beta\) is the FWHM of the peak in radians, and \(\theta\) is the Bragg angle. The analysis indicates that the crystallite size increases with film thickness and is approximately equal to the nominal film thickness, consistent with the trend observed in the rocking curve widths. Thicker films also display higher residual resistance ratios (RRR), with \(\mathrm{RRR} \approx 2.31\), 2.97, and 13.50 for the 5\,nm, 12\,nm, and 21\,nm films, respectively, reflecting enhanced metallic behavior and improved structural uniformity~\cite{roach_nbn_2012}. This trend correlates with higher superconducting transition temperatures ($T_c$) and sharper transitions, with transition widths \(\Delta T\) (defined as the temperature difference between 90\% and 10\% of the normalized resistance) ranging from 0.33\,K to 1.66\,K for the three thicknesses, supporting the above observation~\cite{baeva_natural_2024}. Higher RRR values further indicate improved crystallinity, consistent with complementary structural and transport characterization.

\section{Universal scaling analysis of epitaxial NbN films}

\begin{figure}[ht]
\centering
\includegraphics[width=0.95\linewidth]{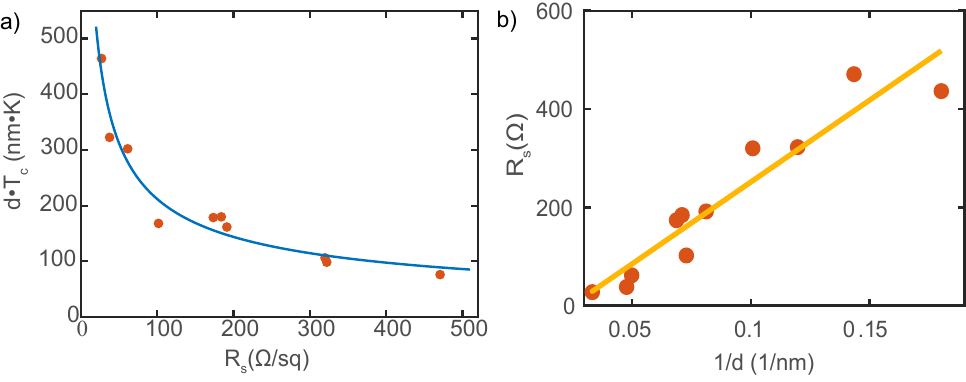}
\caption{\raggedright
\label{fig:millman}
(a) Thickness vs.\ resistance product \( d \cdot T_c \) plotted against sheet resistance \( R_s \), showing an inverse correlation indicative of disorder induced suppression of superconductivity. b) Linear dependence of \( R_s \) on the inverse film thickness \( 1/d \).}
\end{figure}

We analyzed the material properties of epitaxial NbN thin films by testing their consistency with the universal scaling relation proposed by Ivry \textit{et al.}~\cite{ivry_universal_2014}. Figure~\ref{fig:millman}a shows the fit of the measured critical temperature $T_c$, film thickness $d$, and sheet resistance $R_s$ data to the universal scaling law:
\begin{equation}
d T_c = A R_s^{-B}.
\label{eq:ivry}
\end{equation}

This formulation, using the product $d T_c$ rather than $T_c$ alone, captures the coupled influence of dimensionality and disorder on superconductivity. It is well established that $T_c$ decreases as the film becomes thinner and as disorder increases (quantified by the normal-state sheet resistance $R_s$). Thus, plotting $d T_c$ versus $R_s$ provides a more comprehensive description of superconducting trends in ultrathin films than plotting $T_c$ alone.

In our case, the scaling relation fits the experimental data well, yielding $A = 6142.3$ and $B = 0.693$, with a coefficient of determination $R^2 = 0.95$. These parameters are consistent with those reported for epitaxial NbN films grown under comparable conditions~\cite{ivry_universal_2014}, confirming that our samples fall within the same disorder regime and exhibit high structural and superconducting quality. The exponent $B \approx 0.7$ indicates a moderate suppression of superconductivity with increasing disorder, while the prefactor $A$ reflects intrinsic material characteristics such as the clean-limit $T_c$, coherence length, and effective superconducting thickness.

This empirical scaling also provides a practical benefit: $T_c$ can be inferred from normal-state measurements (e.g., $R_s$ at room temperature), offering a convenient metric for assessing film quality and optimizing growth conditions for high-yield SNSPD fabrication.

From a microscopic standpoint, the high $T_c$ of our epitaxial NbN films arises from the close lattice matching between NbN and the sapphire substrate, which minimizes interfacial strain and defect formation in the initial growth layers. This structural coherence promotes robust superconductivity compared with polycrystalline or poorly matched films, where dislocations and strain tend to suppress $T_c$. The Ivry scaling law~\cite{ivry_universal_2014} effectively captures this macroscopic manifestation of disorder: $R_s$ quantifies the degree of scattering, while $d T_c$ reflects the resulting superconducting response. The fitted parameters $A$ and $B$ thus provide a quantitative framework to evaluate how disorder impacts superconductivity in epitaxial NbN.

Finally, Figure~\ref{fig:millman}b shows a linear dependence of the sheet resistance $R_s$ on the inverse film thickness $1/d$, consistent with $R_s = \rho/d$ for films of uniform resistivity $\rho$. This confirms the consistency of resistivity across different thicknesses and supports the validity of applying the Ivry scaling relation to our system. Altogether, the agreement with the Ivry model demonstrates that our deposition conditions yield epitaxial NbN films whose superconducting properties follow universal de-suppression trends and can be predictably tuned via thickness and sheet resistance.

\section{Atomic Force Microscopy}

\begin{figure}[h]
    \centering
    \includegraphics[width=\linewidth]{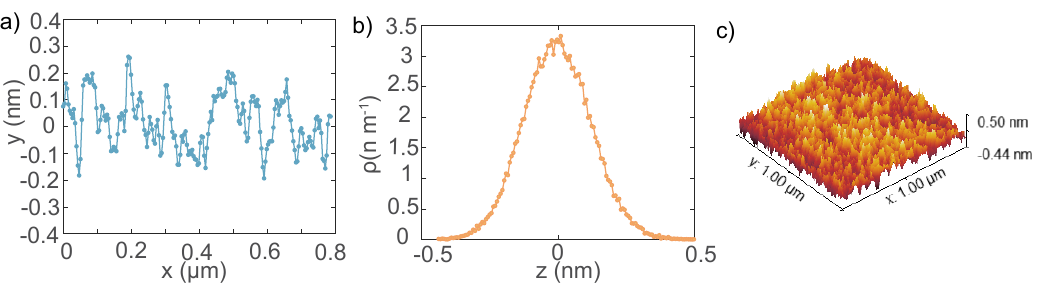} 
    \caption{AFM characterization of epitaxial NbN thin film. (a) Height map (x vs.\ y) showing smooth morphology. (b) Line profile ($\rho$ vs.\ z) indicating minimal surface variation. (c) 3D representation of surface roughness. All data confirm low roughness (~0.4~nm) over a $1~\mu\mathrm{m} \times 1~\mu\mathrm{m}$ scan.}
    \label{fig:S1}
\end{figure}

Atomic force microscopy (AFM) characterization was performed on an 8~nm thick NbN film deposited on sapphire at 800$^\circ$C, with a superconducting critical temperature $T_\mathrm{c} = 11.78$~K and a sheet resistance of 207.41~\textohm/sq. AFM, a surface-sensitive technique, was used to evaluate the surface roughness of the deposited thin film. Over a scanning area of $1~\mu\mathrm{m} \times 1~\mu\mathrm{m}$, the AFM measurement shown in Fig.~\ref{fig:S1} reveals a very low surface roughness of approximately 0.4~nm, which is within the noise level of the instrument. This high degree of surface uniformity is essential for the fabrication of high-performance superconducting devices.

The height map (y vs.\ x) shows a smooth, uniform morphology with no significant particulates or defects, while the line profile ($\rho$ vs.\ z) further confirms minimal surface variation across the scanned area, supporting the structural quality of the epitaxial NbN film.

\section{Electron Backscatter Diffraction}

\begin{figure}[h]
    \centering
    \includegraphics[width=\linewidth]{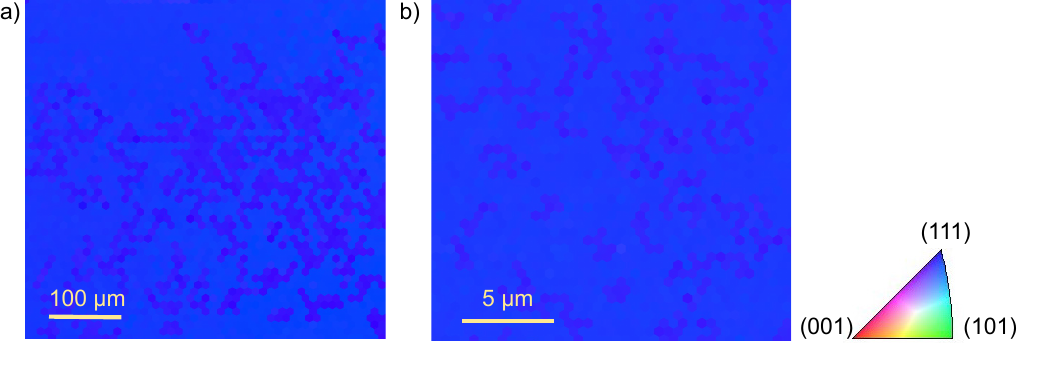} % replace with your AFM image
    \caption{EBSD out-of-plane scans of epitaxial NbN films. Both panels show a strong (111) preferential orientation, confirming high-quality epitaxial growth. (a) Larger area zoom; (b) smaller area zoom.}
    \label{fig:S2}
\end{figure}

Electron backscatter diffraction (EBSD) was performed to confirm the crystallographic orientation and quality of the epitaxial NbN thin films. EBSD provides spatially resolved information on crystal orientation and the presence of grain boundaries, allowing assessment of epitaxial growth and in-plane and out-of-plane alignment.
Fig.~\ref{fig:S2}a and~\ref{fig:S2}b show two different zoom levels of the out-of-plane scan. As expected for films grown on sapphire substrates, the NbN layers exhibit a highly oriented growth in the (111) direction, indicating strong epitaxial alignment. The high degree of crystallographic uniformity confirms the quality of the deposition and suggests minimal formation of misoriented grains or defects within the scanned areas.

\section{X-ray Photoelectron Spectroscopy}

\begin{figure}[h]
    \centering
    \includegraphics[width=\linewidth]{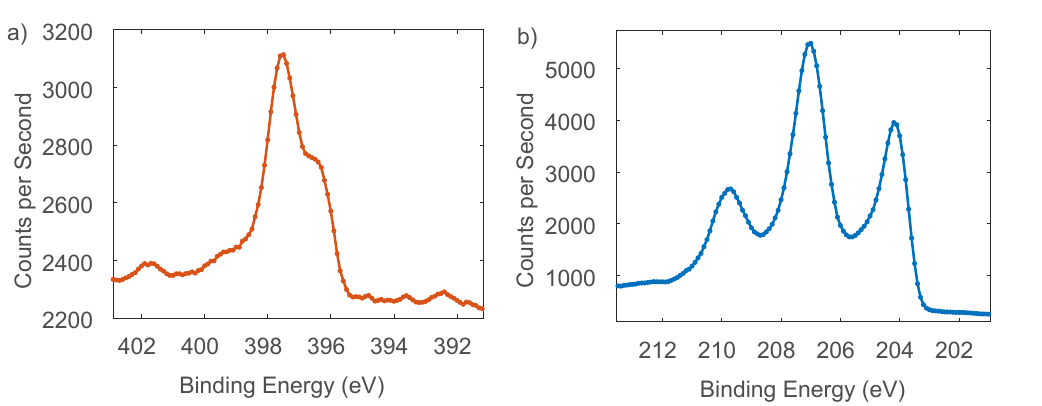}
    \caption{XPS characterization of the NbN film. (a) N~1s spectrum. (b) Nb~3d core-level spectrum showing Nb-N and Nb-O components. Atomic concentrations are summarized in Table~\ref{tab:xps}}
    \label{fig:S3}
\end{figure}

To investigate the surface chemistry of the NbN films, we performed X-ray photoelectron spectroscopy (XPS) on representative samples. The N~1s peak in Fig.~\ref{fig:S3}(a), centered at approximately \SI{397.5}{eV}, corresponds to nitrogen bound in the NbN lattice. The Nb~3d core-level spectrum in Fig.~\ref{fig:S3}(b) shows a dominant doublet centered at \SI{204.2}{eV}, consistent with Nb--N bonds at the surface. Additional components at \SIrange{207}{209.7}{eV} are assigned to Nb--O bonds, reflecting the formation of a native oxide layer upon air exposure.

Quantitative surface analysis yields the following atomic concentrations (corrected by RSF):

\begin{table}[h]
    \centering
    \begin{tabular}{lc}
        \hline
        Element & NbN (\%) \\
        \hline
        C~1s & 30.49 \\
        N~1s & 14.65 \\
        O~1s & 32.57 \\
        Nb~3d & 22.29 \\
        \hline
    \end{tabular}
    \caption{Surface atomic concentrations of the NbN thin film determined from XPS measurements. Carbon is attributed to surface contamination, and oxygen arises from a native oxide layer.}
    \label{tab:xps}
\end{table}

From these surface-sensitive measurements, the Nb:N ratio at the surface is approximately $22.29/14.65 \approx 1.52$. However, this ratio should not be interpreted as the bulk stoichiometry, since XPS probes only the top few nanometers of the film. The significant oxygen and carbon content primarily reflects surface contributions rather than bulk properties.  

\section{Fabrication details}
%\label{sec:SI_fabrication}

Epitaxial NbN thin films were deposited on 10 mm × 10 mm c-cut sapphire substrates by DC reactive ion sputtering. Prior to deposition, the chips were cleaned by sonication sequentially in acetone and isopropyl alcohol (IPA). The base pressure in the deposition chamber was maintained at \(8.6 \times 10^{-9}\) Torr. 

The substrates were mounted on an Inconel sample holder coated with a multilayer of niobium/niobium nitride/titanium. The holder was heated to 800 °C and allowed to stabilize at this temperature for approximately an hour and a half before deposition. The holder was positioned above a Nb target in an Ar and N\(_2\) ambient atmosphere. 

During deposition, the sputtering pressure was held at 2.5 mTorr, with argon and nitrogen flow rates set to 26.5 sccm and 6 sccm, respectively. The sputtering power applied to the target was 400 W (peak power), while a 7 W RF bias was simultaneously applied to the substrate holder to assist in sputter cleaning during the initial target conditioning step~\cite{dane_bias_2017, dane_reactive_2015}. 

After deposition, the samples were cooled down to room temperature under the same pressure conditions, with the heater turned off, and were removed from the chamber once room temperature was reached.

We fabricated superconducting nanowire single-photon detectors (SNSPDs) from 4.5~nm-thick epitaxial NbN films deposited on c-cut sapphire. Nanowires with widths ranging from ultranarrow 20~nm and 30~nm to wider 100~nm devices were patterned using a high-resolution electron-beam lithography process based on 25\% hydrogen silsesquioxane (HSQ)~\cite{yang_fabrication_2005}. The negative-tone resist was spin-coated at 314 rad/s (3000 rpm) for 60~s, resulting in a thickness of approximately 30~nm. Device patterns were defined by electron-beam lithography and developed in 25\% tetramethylammonium hydroxide (TMAH) at room temperature for 2~minutes~\cite{colangelo_superconducting_2023}. The resulting patterns were then transferred into the NbN films using reactive ion etching with CF$_4$. Nanowire widths were verified by scanning electron microscopy (SEM) on the HSQ mask after deposition of a 2~nm gold layer, ensuring precise control of the nanowire dimensions.

\begin{figure}[h]
    \centering
    \includegraphics[width=\linewidth]{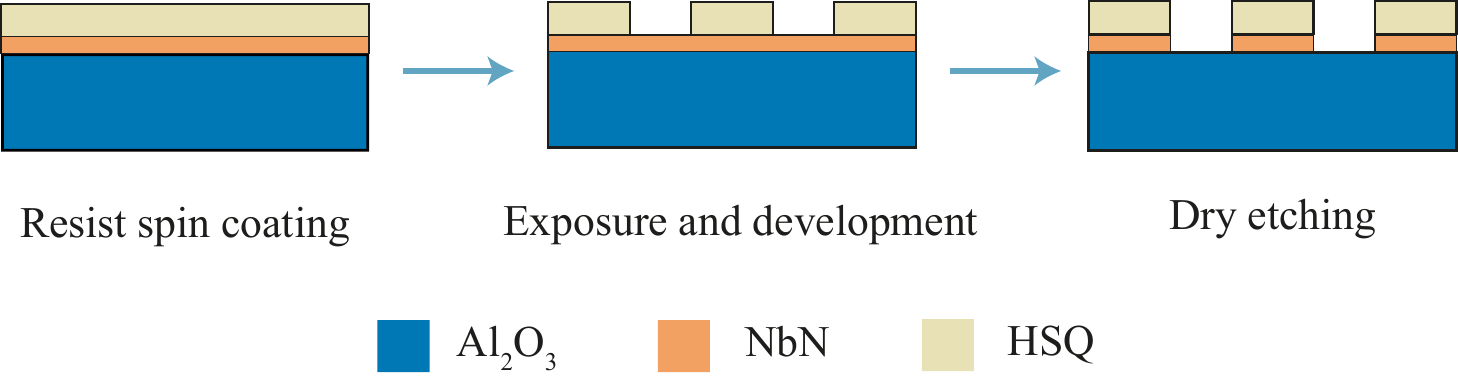} 
    \caption{Schematic of the fabrication process for ultranarrow NbN nanowire devices on epitaxial thin films. This workflow enables precise pattern transfer and uniform nanowire widths below 2~nm variation.}
    \label{fig:S4}
\end{figure}

\section{Measurements Methods}

\begin{figure}[H]
    \centering
    \includegraphics[width=\linewidth]{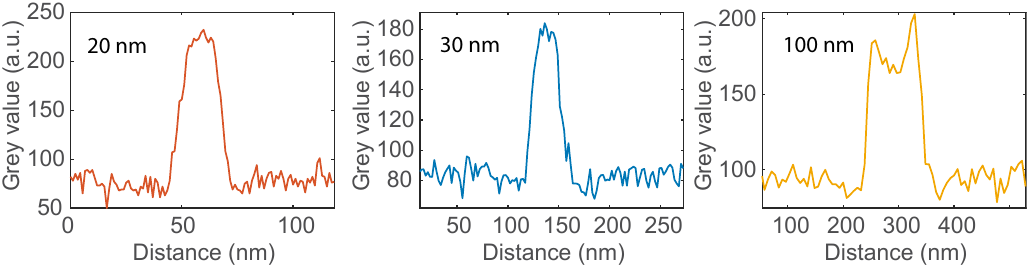} 
    \caption{Nanowire widths extracted from SEM images. Grey value line profiles perpendicular to each nanowire were analyzed in ImageJ to identify edge transitions. The width, defined as the distance between edges, was measured across multiple profiles per device, confirming nominal widths of 100~nm, 30~nm, and 20~nm}
    \label{fig:widths}
\end{figure}

X-ray diffraction (XRD) measurements were performed to assess the crystallographic quality of the NbN films. High-resolution $\theta$--2$\theta$ scans were acquired using a Rigaku SmartLab diffractometer operated at 40\,kV and 30\,mA with Cu K$\alpha$ radiation ($\lambda = 1.5406$\,\AA). Azimuthal ($\phi$) scans were carried out using a PANalytical X’Pert PRO system operated at 30\,kV and 10\,mA, using a sealed tube with a copper target, producing Cu K$\alpha$ radiation ($\lambda = 1.5406$\,\AA), to evaluate the in-plane epitaxial relationships and detect the presence of twin domains.
The critical temperature ($T_c$) of the devices was determined by measuring the resistance as a function of temperature during both warming and cooling cycles. The temperature was controlled using a cryogenic temperature controller with a standard PID feedback loop, based on readings from a calibrated low-temperature thermometer positioned in close thermal contact with the sample stages.
Nanowire widths were estimated from SEM images by analyzing the grey value intensity across each nanowire using ImageJ. Line profiles were drawn perpendicular to the nanowire axis to extract the grey value as a function of distance (Fig.~\ref{fig:widths}). The nanowire edges were identified from the transitions in the intensity profiles, and the width was taken as the distance between these edges. Multiple line profiles were analyzed per device to obtain representative widths. Using this method, the nominal nanowire widths of 100~nm, 30~nm, and 20~nm were confirmed with high fidelity.

\begin{figure}[H]
    \centering
    \includegraphics[width=0.7\linewidth]{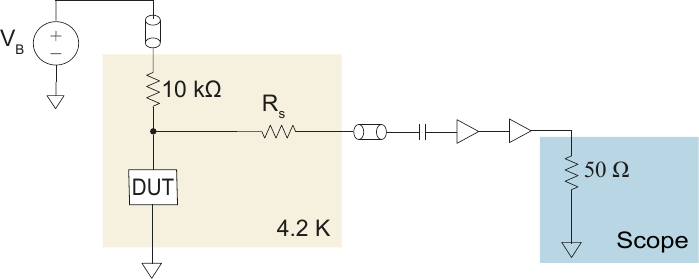} 
    \caption{Schematic of the readout electronics setup for the nanowire detector metrics measurements.}
    \label{fig:circuit}
\end{figure}

The electronic properties of the nanowires were measured using a custom-built dip probe immersed in liquid helium~\cite{butters_digital_2022}, operating at a temperature of 4.2\,K inside a helium dewar. The readout electronics setup consists of a voltage source in series with a bias resistor: a 10\,k$\Omega$ resistor was used for the 100\,nm wide nanowire, while 100\,k$\Omega$ resistors were used for the 20\,nm and 30\,nm devices to ensure stable biasing at lower switching currents (Fig.~\ref{fig:circuit}). The output pulses generated by photon detection events are AC-coupled via a capacitor to the amplification and acquisition chain. The SNSPDs were optically tested at two key wavelengths, 1550~nm and 780~nm through a fiber-optic cable.
For reset time measurements, a surface-mount resistor of 215~$\Omega$ ($R_{s}$) was added in series with the 50~$\Omega$ amplifier input on the PCB to increase the load resistance and reduce the $L/R_{\mathrm{load}}$ time constant, where $R_{\mathrm{load}} = R_{s} + 50~\Omega$. The device kinetic inductance was estimated as $L = 1.38\, (R_{\mathrm{sheet}} / T_c ) \, (L/W)$, where $R_{\mathrm{sheet}}$ is the sheet resistance of the superconducting film, $T_c$ its critical temperature, and $L/W$ is the number of squares~\cite{colangelo_superconducting_2023}. This configuration effectively shortens the detector recovery time.

\section{Detector Performance}

\begin{figure}[h!]
    \centering
    \includegraphics[width=\linewidth]{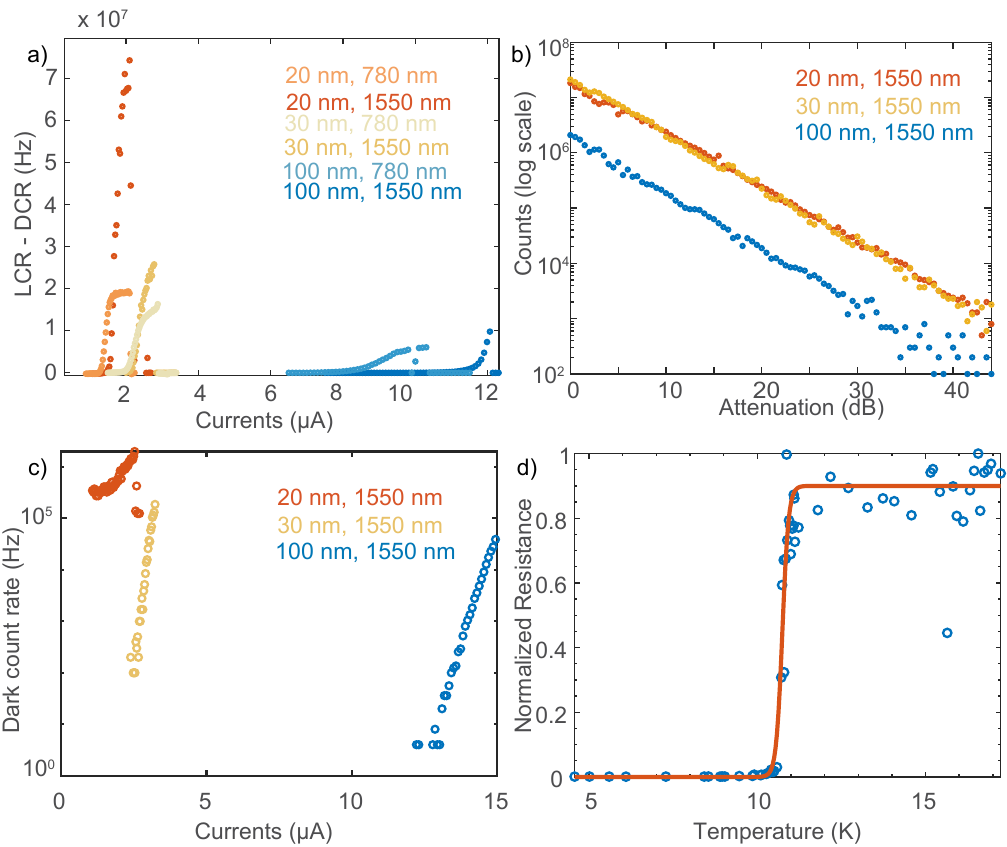} 
    \caption{ (a) Photon count rate (Light counts - Dark counts) as a function of bias current for NbN nanowires of different widths (20~nm, 30~nm, and 100~nm) measured at 4.2~K. (b) Linearity check of NbN nanowire devices. Count rate is plotted as a function of incident photon flux for nanowires of different widths (20~nm, 30~nm, and 100~nm). (c) Dark count rates as a function of bias current for NbN nanowires of different widths (20~nm, 30~nm, and 100~nm) measured at 4.2~K. (d) Superconducting transition temperature ($T_c$) for 20 nm wide NbN nanowires. The $T_c \approx 10.75$~K. The transition curves were fitted using a sigmoidal function to accurately extract $T_c$.}
    \label{fig:S6}
\end{figure}
We characterized the photon count rate (PCR) of NbN nanowire devices with different widths (20~nm, 30~nm, and 100~nm) at an operating temperature of 4.2~K (Fig.~\ref{fig:S6}(a)). The PCR curves, obtained under continuous-wave laser illumination at 1550~nm and 780~nm wavelengths with a laser power of 0.5~mW, show the dependence of detection efficiency on the bias current. For the 20~nm-wide nanowires, we observe the highest saturation plateau, corresponding to the maximum count rate, while wider wires (30~nm and 100~nm) display progressively lower detection efficiencies. We also extracted the switching currents ($I_\text{sw}$) of the devices. As expected, $I_\text{sw}$ scales with the nanowire cross-sectional area, narrower nanowires (20~nm) exhibit lower switching currents compared to wider ones (100~nm).

To verify that the counts correspond to single-photon detection events, we measured the linearity of the count rate as a function of the incident photon flux  (Fig.~\ref{fig:S6}b). The linear dependence confirms the single-photon detection regime for all device widths.

Moreover, Fig.~\ref{fig:S6}c shows the dark count rate (DCR) as a function of bias current for NbN nanowires of different widths measured at 4.2 K. The 20~nm-wide nanowires exhibit a markedly higher dark count rate and a steeper dependence on bias current compared to the 30~nm and 100~nm devices. This behavior suggests enhanced edge sensitivity, likely arising from increased current crowding and local nonuniformities at reduced dimensions.

Finally, we measured the superconducting transition temperature ($T_c$) of the nanowires (Fig.~\ref{fig:S6}d) using a bias current of 1~µA. All devices exhibit sharp transitions with $T_c \approx 10.5$~K, consistent across different wire widths. For the 20-nm-wide wire, we extracted $T_c = 10.746$~K with a narrow transition width of $\Delta T = 0.092$~K. The data were fitted with a sigmoidal function to determine $T_c$ at 50\% of the transition.

\section{Temporal response and exponential fits of the detector}

\begin{figure}[h]
    \centering
    \includegraphics[width=0.5\linewidth]{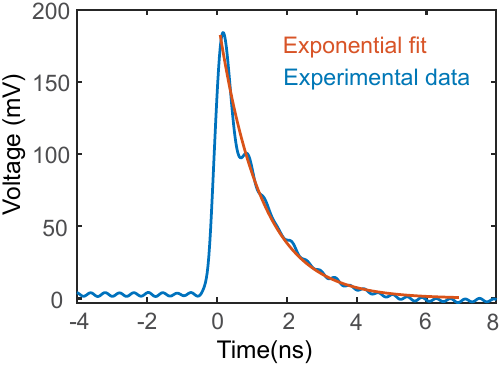}
    \caption{Temporal response of a superconducting nanowire single-photon detector (SNSPD) with and without an additional load resistor. Pulse trace of the detector without the load resistor, showing a long exponential decay with a time constant of 1.2~ns.}
    \label{fig:pulse}
\end{figure}

In a simple switch model for the SNSPD, after the wire returns superconducting, the current recovers with a time constant $\tau$ =  $L/R_{\mathrm{load}}$, where L is the inductance of the SNSPD plus any parasitic inductance and $R_{\mathrm{load}} = R_{s} + 50~\Omega$. The voltage measured by the scope decays accordingly with the same time constant.
Fig.~\ref{fig:pulse}a shows the temporal response of one of the measured superconducting nanowire detector without an additional load resistor. The pulse exhibits a long decay tail that is well described by an exponential fit, from which a decay time constant of 1.2~ns is extracted. 

\begin{figure}[h]
    \centering
    \includegraphics[width=0.5\linewidth]{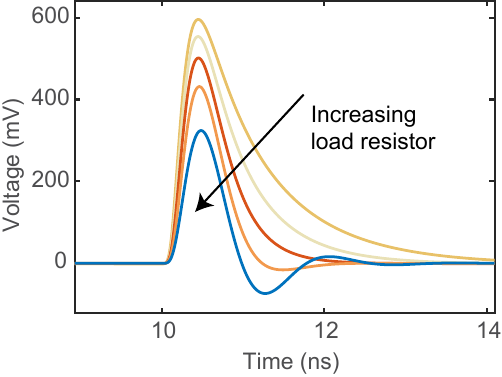}
    \caption{SPICE simulation of the transient response of our circuit for five different series resistance values. Increasing the series resistance decreases the effective $L/R$ time constant, resulting in a faster decay of the output pulse.}
    \label{fig:spice}
\end{figure}

To investigate the impact of the load on the detector dynamics, several measurements were performed with different load resistances connected in series with the device. Increasing the load resistance progressively shortened the decay time, consistent with the expected reduction of the effective $L/R$ constant. 
To support this interpretation, we also performed LTspice simulations of the circuit shown in Fig.~\ref{fig:circuit}, which confirm that adding the load resistor reduces the pulse width and decay time. 
The simulation reproduces the experimental setup in which an ideal voltage source in series with a resistance provides a current bias for the SNSPD, modeled using a dynamic LTSpice model~\cite{berggren_superconducting_2018}. The bias resistor $R_\mathrm{s}$ is connected to a bias tee, with the DC port shunted to ground through a $50~\Omega$ termination. The AC port is connected to a chain of low-noise amplifiers (LNAs), which in turn feed a $50~\Omega$ load representing the oscilloscope input. To account for pulse distortion due to frequency cutoff effects, the LNA and bias tee models include their respective low- and high-frequency cutoffs. The bias tee is a model ZFBT-4R2GW+ from Mini-Circuits, with a high cutoff frequency of $f_\mathrm{H} = 4.2~\mathrm{GHz}$. The first amplifier is a LNA-2500 from RFBay Inc., with $f_\mathrm{H} = 2.5~\mathrm{GHz}$, while the second amplifier is a LNA-2000 from RFBay Inc., with $f_\mathrm{H} = 2.0~\mathrm{GHz}$. The oscilloscope high-frequency cutoff of $6~\mathrm{GHz}$ was not modeled, as it exceeds the bandwidth of the preceding amplifiers. A $1~\mathrm{pF}$ parasitic capacitance to ground was included between the SNSPD source and the load resistor. This value was chosen as a reasonable estimate based on the PCB-mounted load resistor and the wire-bond connections from the device to the PCB pads. The parasitic series inductance due to the wire bonds was modeled using inductors with $L_\mathrm{s} = 5~\mathrm{nH}$. The bias current is $19~\mu\mathrm{A}$, the measured sheet resistance is $R_\square = 491~\Omega$, the measured critical temperature is $T_\mathrm{c} = 10.5~\mathrm{K}$, the length of the wire is $2~\mu\mathrm{m}$ and its width is $100~\mathrm{nm}$, the film thickness is $5.3~\mathrm{nm}$, and the total number of squares in the superconducting layer is $n_\mathrm{s} = 500$. The calculated sheet inductance is $L = 1.38 \,(R_\square/T_\mathrm{c})\, n_\mathrm{s} \times 10^{-12}~\mathrm{H} \simeq 32.3~\mathrm{nH}$. Voltage traces were rescaled to match the experimental amplitudes.
Device pulses were simulated while sweeping the load resistance value. We reproduced the experimental observation that increasing the resistive load decreases the pulse width, confirming the role of $L/R$ reduction in achieving faster detector electrical reset (Fig.~\ref{fig:spice}). When including a $1~\mathrm{pF}$ parasitic capacitor between the SNSPD drain and the load resistor, however, increasing $R_s$ causes the simple $L/R$ recovery model to fail to accurately describe the dynamics. Once the capacitive time constant $RC$ becomes comparable to the inductive time constant $L/R$, the early-time dynamics are no longer dominated purely by inductive decay. When $RC > 4L/R$, the parallel RLC response becomes underdamped, leading to the slight post-pulse ringing observed in the measured voltage traces for $R_s = 215~\Omega$. Load resistance values ranging from 215 $\Omega$ to 0~$\Omega$ were simulated. Distributed effects, including the non-linear transmission line behavior of the superconducting nanowire and reflections from non-matched tapers, have not been accounted for.

\section{Nanowire-to-substrate thermal conductance}

\begin{figure}[h]
    \centering
    \includegraphics[width=0.5\linewidth]{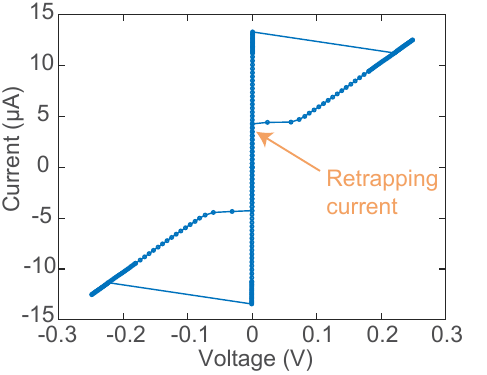}
    \caption{Current-voltage characteristics of a 100~nm-wide NbN nanowire measured at 4.3~K. The retrapping current ($I_r$) is indicated, marking the transition from the resistive to the superconducting state during the decreasing current sweep.}
    \label{fig:S8}
\end{figure}

We estimated the thermal conductance between the NbN nanowires and the substrate using the retrapping and switching currents measured at 4.3~K. The nanowire-to-substrate thermal conductance, $h_c$, can be expressed as~\cite{berggren_superconducting_2018}:

\begin{equation}
h_c = \frac{R_{sq} \, I_r^2}{2 \, w^2 \, (T_c - T_s)},
\end{equation}

where $R_{sq}$ is the sheet resistance of the NbN film, $I_r$ is the retrapping current estarcted from a current-voltage characteristic (Fig.~\ref{fig:S8}), $w$ is the nanowire width, $T_c$ is the superconducting critical temperature of the wire, and $T_s$ is the substrate temperature.

The resulting nanowire-to-substrate thermal conductance is:

\[
h_c^{\rm epitaxial} \approx 9.3 \times 10^4~{\rm W\,m^{-2}\,K^{-1}}.
\]

For comparison, thermal conductance on SiO$_2$ (thermal oxide) is significantly lower, on the order of $10^2$–$10^3~{\rm W\,m^{-2}\,K^{-1}}$~\cite{giri_effect_2016}. This shows that Al$_2$O$_3$ substrates provide much better thermal coupling, allowing faster thermal relaxation and recovery of the nanowire after photon detection.

\section{Magnetotransport characterization}

\begin{figure}[ht]
\centering
\includegraphics[width=\linewidth]{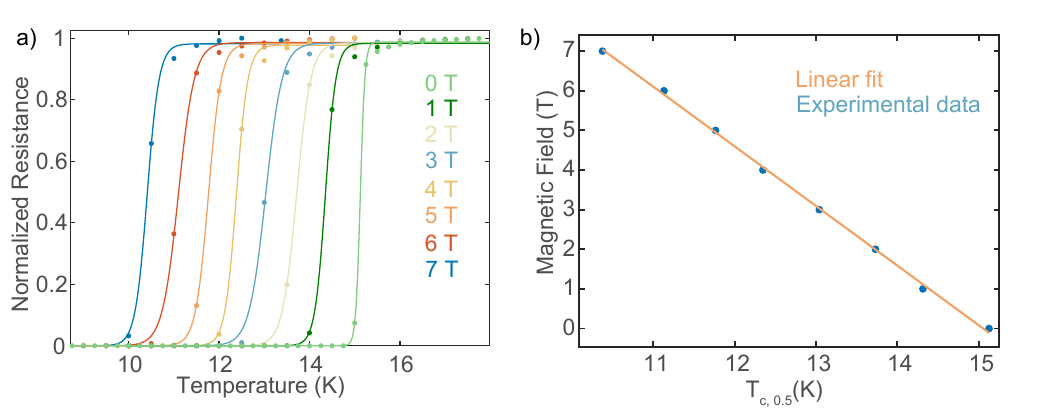}
\caption{\raggedright
\label{fig:mpms}
(a) Normalized resistance as a function of temperature for an 12~nm NbN thin film measured under different perpendicular magnetic fields (0–7~T). Each dataset (circles) represents the resistance normalized to its maximum value below 18~K, obtained via four-probe measurements. A sigmoidal fit (solid lines) is applied to extract the superconducting critical temperature \(T_c\). 
%The superconducting transition systematically shifts to lower temperatures and broadens with increasing magnetic field, consistent with the suppression of superconductivity in the presence of an external field. 
(b) Magnetic field dependence of the superconducting transition temperature \( T_{c,0.5} \), defined as the temperature at which the resistance drops to 50\% of its normal-state value. The data points (circles) are extracted from the fits in Figure~\ref{fig:mpms}a. A linear fit (solid line) is applied.
}
\end{figure}

The upper critical field \(H_{c2}(T)\) as a function of temperature was determined using a Quantum Design Magnetic Property Measurement System (MPMS) system by applying magnetic fields up to 7\,T perpendicular to the film surface (\(H \perp S\)). Fig.~\ref{fig:mpms} shows the critical temperature $T_c$ under different perpendicular magnetic fields for a 12~nm. The superconducting transition shifts to lower temperatures and broadens with increasing field.
Each transition curve was fitted using a sigmoidal function to extract $T_{c,0.5}$ at 50\%. Fig.~\ref{fig:mpms}b displays $T_{c,0.5}$ as a function of magnetic field, showing the clear linear dependence.
For the 12~nm, the upper critical field at zero temperature, $\mu_0 H_{c2}(0)$, is estimated using the Werthamer–Helfand–Hohenberg (WHH)~\cite{werthamer_temperature_1966} approximation in still in the dirty limit as $\mu_0 H_{c2}(0) = 0.69\, T_c \left| \frac{dH_{c2}}{dT} \right|_{T \to T_c} = 15.7~\mathrm{T}$. The superconducting coherence length is $\xi(0) = \sqrt{\frac{\Phi_0}{2 \pi B_{c2}(0)}} = 5.74~\mathrm{nm}$, and the electron diffusivity is $D = \frac{4 k_B}{\pi e} \cdot \left| \frac{dB_{c2}}{dT} \right|^{-1} = 0.73~\mathrm{cm^2/s}$. The higher values observed for the thicker 12~nm film are consistent with stronger scattering in the thinner 5~nm films.

\section{Ab-Initio Modeling}

To model the photon count rate (PCR) behavior and predict the saturation current of the epitaxial SNSPDs, we used a multiscale theoretical framework combining \textit{ab initio} kinetic modeling with time-dependent Ginzburg–Landau (TDGL) simulations~\cite{simon_ab_2025}. This approach captures the sequence of nonequilibrium processes triggered by photon absorption in the superconducting nanowire. The early stages of detection are described using an \textit{ab initio} kinetic equation formalism operating on sub-picosecond timescales, which models the energy partitioning between quasiparticles and phonons and accounts for phonon escape into the substrate. The resulting non-equilibrium quasiparticle distributions are then used as input for TDGL simulations, which evolve the superconducting order parameter on tens-of-picoseconds timescales. This enables a direct prediction of whether the absorbed photon drives the system into the resistive state, producing a detectable voltage pulse. 
The combined framework provides a binary prediction of photon detection under specific bias conditions. We employ our previously developed \textit{ab initio} model~\cite{simon_ab_2025}, which requires no adjustable parameters to obtain the saturation current, and only a fit to the fluctuation rate during the down-conversion of photon energy to reproduce the experimental PCR dependence. When fitting, we assume that the normalized PCR distribution is well-represented by a sigmoid centered at the saturation current and normalized to one. The agreement between theory and experiment confirms that the multiscale model reliably captures the detection dynamics of our epitaxial NbN SNSPDs, giving a valuable feedback loop between modeling and experiment for device optimization.

%\bibliographystyle{unsrt}
%\bibliography{references}

\printbibliography
%\bibliography{acs-template.bib}
\end{document}